# A new poling protocol for enhanced piezoelectricity in $Bi_{0.5}Na_{0.5}TiO_3$

Mukesh Kumari[1,2], Ratnamala Chatterjee[1a)]

[1]Magnetics & Advanced Ceramics Laboratory, Physics Department, Indian Institute of Technology Delhi 110016, India

[2]Department of Materials Science and Engineering, University of California, Berkeley, CA 94720, USA

In this work, a way to improve the piezoelectric properties of $Bi_{0.5}Na_{0.5}TiO_3$ (BNT) is demonstrated by introducing a new poling protocol. A customized corona poling unit with a low temperature (~77 K) sample stage is suggested. Using this protocol, the BNT sample is quenched from its paraelectric phase ($T$~350°C) directly to its ferroelectric phase range ($T$< 180°C) under corona discharge. Sample poled under this protocol showed an immense improvement (~38% increase) in the piezoelectric coefficient ($d_{33}$) and ~ 20% increase in the maximum unipolar piezoelectric strain ($S_{max}$%).

[a)]Electronic mail: ratnamalac@gmail.com, rmala@physics.iitd.ac.in,

Lead-based piezoelectric ceramics exemplified by Pb(Zr,Ti)O$_3$ (PZT) are widely used for sensors, actuators, and ultrasonic motors owing to their excellent piezoelectric properties.[1-4] However, the use of these ceramics has caused serious environmental problems and proved harmful to human health and environment due to the strong toxicity of Pb. Therefore, intensive efforts have been continuously devoted to develop lead-free/environment-friendly piezoelectrics with properties comparable to the Pb-based ceramics.[5-8] Among the noted lead-free piezoelectrics, Bi$_{0.5}$Na$_{0.5}$TiO$_3$ (BNT) and its solid solutions with other perovskite oxides found their place as probable replacement for Pb-based MEMS device applications.[9-11] Pure BNT is a rhombohedral ferroelectric with perovskite structure having large remnant polarization $P_r$ ~38 μC/cm$^2$ comparable to that of PZT ($P_r$ ~31 μC/cm$^2$) and high Curie temperature $T_c$ ~350°C ($T_c$ ~ 386°C for PZT)[12-14]. The notable difference between the two systems is that for PZT, tetragonal ferroelectric (T-FE)$_{PZT}$ phase changes to cubic paraelectric (C-PE)$_{PZT}$ phase at ~386 °C, whereas in BNT, the rhombohedral ferroelectric (R-FE)$_{BNT}$ phase does not directly change to cubic paraelectric (C-PE)$_{BNT}$ phase, instead an additional tetragonal antiferroelectric (T-AFE)$_{BNT}$ phase in the temperature range ~180°C-350°C exists. Thus, for BNT, the sequence goes as, Rhombohedral FE – Tetragonal AFE – Cubic PE.[12,15-18]

In comparison to Pb-based ceramics, pure BNT is known to show reasonably low piezoelectric coefficient ($d_{33}$) ~80 pC/N ($d_{33}$ >250 pC/N for PZT) and large coercive field ($E_c$) ~ 70 kV/cm ($E_c$~ 20 kV/cm for PZT)[1,12]. These two practical parameters restrict the use of this otherwise attractive ferroelectric non-Pb-based composition, BNT. In literature, the low values of $d_{33}$ in BNT have been attributed to its large $E_c$ and difficulty in poling process.[12,14]

In this letter, we are demonstrating that while poling BNT ceramics, if one can quench the AFE phase region (180 °C- 350 °C), an immense improvement can be achieved in the piezo response of this hard ferroelectrics.

Polycrystalline samples of $Bi_{0.5}Na_{0.5}TiO_3$ were prepared by conventional solid state reaction route using the high purity (≥ 99% purity Sigma Aldrich) $Bi_2O_3$, $Na_2CO_3$ and $TiO_2$ powders. The powders were weighed according to stoichiometric formula and ball milled for 24 h in acetone medium. This mixture is then dried and calcined at 950 °C for 12 h. Calcined powders were pulverized and mixed with polyvinyl alcohol (PVA) used as binder and then pressed into pellets of 10 mm diameter. The sintering of pellets was carried out at 1150 °C for 2 h.

The phase formation and purity of thus prepared sintered BNT pellet was tested using X-ray diffraction (CuKα radiation, λ = 1.54178 Å) (Philips X-pert PRO) as shown in Fig.1. A well crystallized and pure single perovskite phase is obtained with no extra peak belonging to any impurity phase. The peaks are indexed as rhombohedral crystal system using X-Powder software and matches well with the literature.[12]

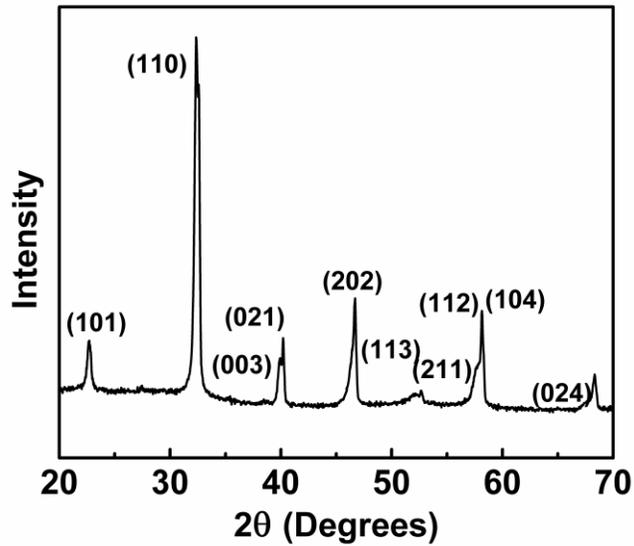

**Fig. 1:** Room temperature X-ray diffraction pattern of sintered pellet of BNT.

The ferroelectric hysteresis (*i.e.* polarization (P) vs. electric field (E)) loops at room temperature were measured at 1Hz using a ferroelectric tester (Radiant Precision Premier II Technology). As reported earlier by other researches as well[12-14], the sample shows a well saturated ferroelectric loop (shown in Fig. 2) with remanent polarization $P_r$ ~38 µC/cm$^2$ with a high coercive electric field $E_c$ ~65 kV/cm.

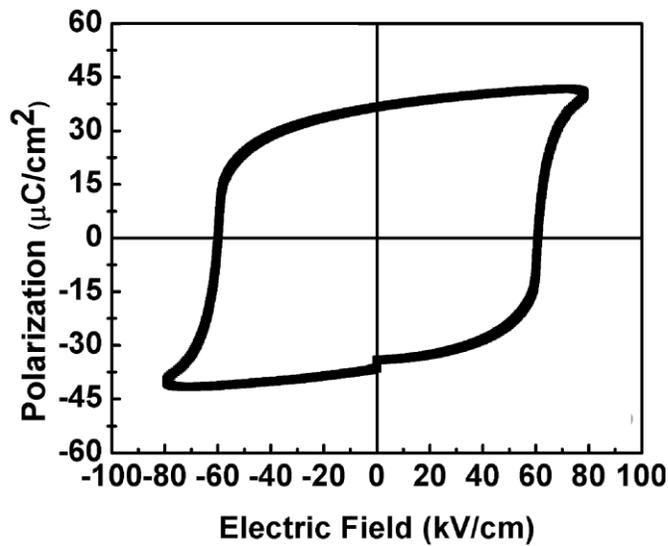

**Fig. 2:** Polarization versus Electric Field for BNT sample at room temperature.

The results of dielectric measurements agree well with the best values for BNT reported in literature.[12,19] Figure 3 shows the dielectric data depicting clearly that the sample goes from rhombohedral ferroelectric (R-FE)$_{BNT}$ phase to tetragonal antiferroelectric (T-AFE)$_{BNT}$ phase at $T_d$ (depolarization temperature) ~180 °C and the peak at higher temperature $T_m$ (maximum temperature) ~350 °C evidences the tetrahedral antiferroelectric (T-AFE)$_{BNT}$ phase to cubic paraelectric (C-PE)$_{BNT}$ phase transition.

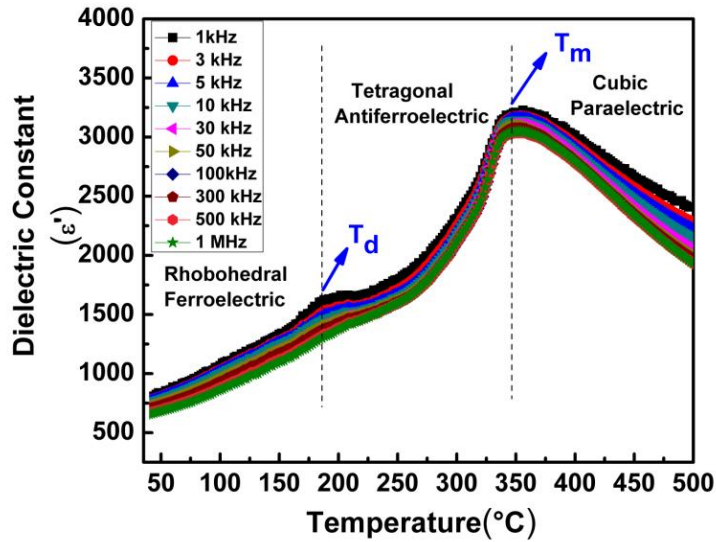

**Fig. 3:** Dielectric Constant for BNT sample at varying frequencies.

The piezoelectric charge coefficient ($d_{33}$) of BNT was measured using YE 2730A d$_{33}$ meter, after poling the sample electrically. In case of conventional poling, the sample is being heated at a temperature above that of the ferroelectric Curie temperature (~350°C in case of BNT) for half an hour so that all the dipoles becomes random (paraelectric state). Then an electric field is applied (along with the heating ON) for another half an hour, so that corona discharge is formed. Switching off heater and cooling down the sample in the presence of applied electric field for another half an hour completes the poling process. The measured value of $d_{33}$ was found ~80 pC/N, again, in accordance with the best values reported in the literature.[12,20-21]

At temperature $T > T_m$, while the sample is paraelectric and the dipoles are randomly oriented, it is customary to apply high voltages to orient these dipoles for obtaining best $d_{33}$ values. However, in our BNT sample, as we apply large voltages at temperature $T > T_m$ when the dipoles are random; on cooling, the dipoles have to first enter into AFE phase (~ 350°C-180°C). All the dipoles will try to align antiparallel to each other and then on further cooling below $T < T_d$, as the sample enters into FE phase range, they will start orienting parallel to the field direction. However, here the BNT in between passes through AFE region, where the dipoles had been arranged anti-parallely, it requires more energy to orient them all completely in the direction of the field. This leads to poor $d_{33}$ values in BNT based ceramics.

In order to avoid the AFE region, in this proposed new protocol, we have quenched our sample from 350 °C to 77 K under corona discharge. For this process, we have customized our corona poling unit as shown in the Fig. 4. In this protocol, first, the sample is being heated to a temperature of ~350°C (without electric field) for half an hour and then with electric field (corona discharge) for another half an hour. After that, we quenched the sample by switching off the heater and simultaneously, replacing this sample stage with a liquid nitrogen temperature sample stage such that the sample does not have time to enter in antiferroelectric region. This way, sample is being cooled in presence of corona for another half an hour. After the poling, the measured piezoelectric coefficient was found to be ~110 pC/N. Thus, the poling of BNT based oxides following this new poling protocol leads to an immense improvement (~38% increase in BNT sample) in piezoelectric coefficient[21].

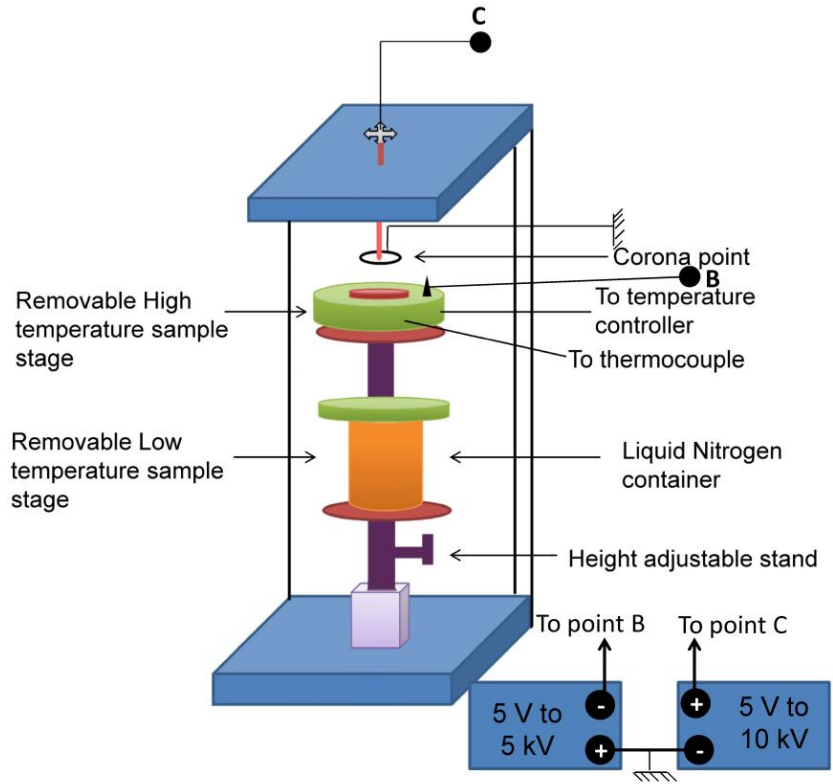

**Fig. 4:** Schematic diagram of customized corona poling unit.

Figures 5(a)-5(d) show the unipolar and bipolar piezoelectric strain loops respectively, measured on the BNT sample poled conventionally (Figs. 5(a) and 5(b)) and using new poling protocol (Figs. 5(c) and 5(d)). Large increase in both unipolar (~ 20%) and bipolar (~25%) maximum piezoelectric strain % ($S_{max}$%) is observed in the BNT sample poled using new protocol.

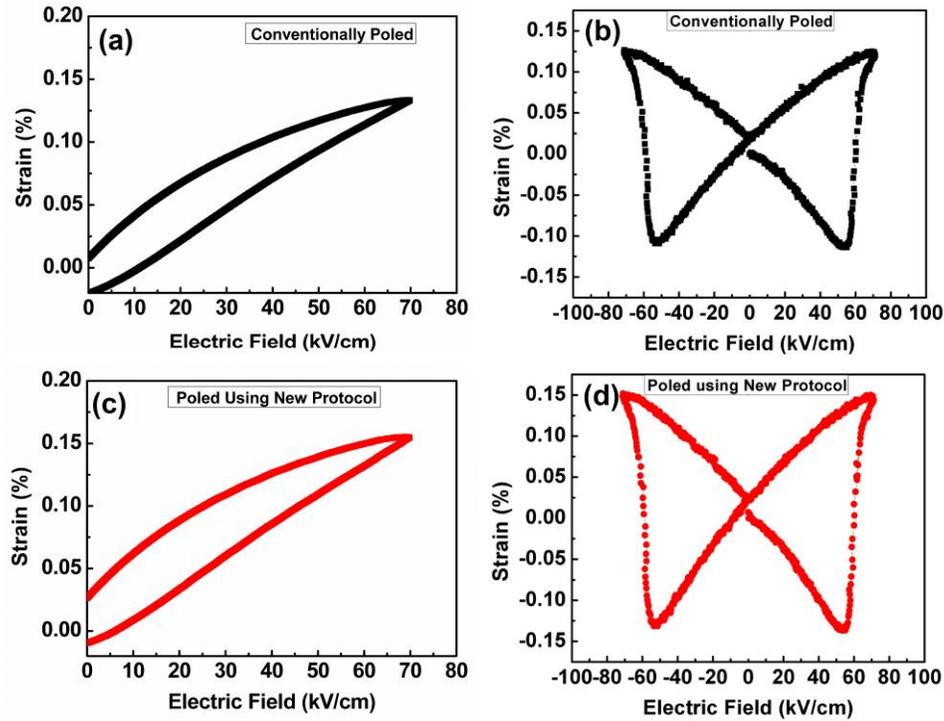

**Fig. 5:** (a) Unipolar piezoelectric strain loop (b) Bipolar piezoelectric strain loop; for BNT sample poled conventionally (c) Unipolar piezoelectric strain loop (d) Bipolar piezoelectric strain loop; for BNT sample poled using new poling protocol.

In conclusion, we are proposing a new poling protocol to pole BNT based oxides. A new customized corona poling unit has been suggested and used for this purpose. The piezoelectric coefficient of BNT shows ~38% increase when poled using this new protocol poled. Similarly, the maximum unipolar piezoelectric strain ($S_{max}$%) is found increased by ~20%.

## ACKNOWLEDGMENTS

One of the authors (M.K) would like to acknowledge DST, India for providing the fellowship. The authors would also like to acknowledge DRDO, India for financial support.

# References

[1] Takashi Yamamoto, Jpn. J. Appl. Phys. **35**, 5104 (1996).

[2] B. Jaffe, W. R. Cook, Jr., and H. Jaffe: Piezoelectric Ceramics (Academic Press, London, 1971).

[3] K. Uchino, Solid State Ionics 108, 43 (1998)

[4] K. Uchino, Ferroelectric Devices (Marcel Dekker, New York, 2000).

[5] Jürgen Rödel *et.al.*, J. Eur. Ceram. Soc. **35,** 1659 (2015).

[6] Wenping Cao *et.al.*, Appl. Phys. Lett. **108**, 202902 (2016).

[7] J. Zhang, et al. Nat. Commun. **6:6615**, 7615 (2015).

[8] J. Rödel *et.al.*, J. Am. Ceram. Soc., **92**, 1153 (2009).

[9] E. Birks *et.al.*, J. Appl. Phys. **119**, 074102 (2016).

[10] Pedro B.Groszewicz *et.al.*, Sci. Rep., **6**, 31739 (2016).

[11] Y. S. Sung *et.al.*, App. Phys. Lett., **98**, 012902 (2011).

[12] A. Singh and R. Chatterjee, J. Appl. Phys., **109**, 024105 (2011).

[13] Singh and R. Chatterjee, J. Am. Ceram. Soc. **96**, 509 (2013).

[14] Y. R. Zhang, J. F. Li, B. P. Zhang, and C. Peng, J. Appl. Phys., 103, 074109 (2008).

[15] Wei Peng, *et.al.* Appl. Phys. Lett. **106**, 092903 (2015).

[16] Qian Gou *et.al.*, RSC Adv., **5**, 30660 (2015).

[17] Hongfeng Lu, Shanying Wang, and Xiaosu Wang, J. Appl. Phys. **115**, 124107 (2014).

[18] Yoshinori Watanabe, Yuji Hiruma, Hajime Nagata, and Tadashi Takenaka, Ferroelectrics, **358**, 139 (2007).

[19] Hiroki Muramatsu, Hajime Nagata, and Tadashi Takenaka, Japanese Jpn. J. Appl. Phys. **55**, 10TB07 (2016).

[20]Yoshinori Watanabe, Yuji Hiruma, Hajime Nagata and Tadashi Takenaka, Key Eng. Mater. **388**, 229 (2008).

[21]Yuji Hiruma, Hajime Nagata, and Tadashi Takenaka, J. Appl. Phys. **105**, 084112 (2009).

**Figure Captions**

**Fig. 1:** Room temperature X-ray diffraction pattern of sintered pellet of BNT.

**Fig. 2:** Polarization versus Electric Field for BNT sample at room temperature.

**Fig. 3:** Dielectric Constant for BNT sample at varying frequencies.

**Fig. 4:** Schematic diagram of customized corona poling unit.

**Fig. 5:** (a) Unipolar piezoelectric strain loop (b) Bipolar piezoelectric strain loop; for BNT sample poled conventionally (c) Unipolar piezoelectric strain loop (d) Bipolar piezoelectric strain loop; for BNT sample poled using new poling protocol.